\documentstyle[12pt,preprint]{aastex}

\newcommand {\hii}{H\,{\sc ii}} 
\newcommand {\kms}{\relax \ifmmode {\,\rm km\,s}^{-1}\else \,km\,s$^{-1}$\fi}
\newcommand {\ha}{H$\alpha$}
\newcommand {\oiii}{[O\,{\sc iii}]}
\newcommand {\nii}{[N\,{\sc ii}]}

\shorttitle{Bubbles of Massives stars}
\shortauthors{Naz\'e et al.}

\begin{document}

\title{Interstellar Bubbles in Two Young \hii\ Regions}

\author{Ya\"el Naz\'e\altaffilmark{1,}\altaffilmark{2,}\altaffilmark{5},
  You-Hua Chu\altaffilmark{1,}\altaffilmark{6},
  Sean D.\ Points\altaffilmark{1,}\altaffilmark{6}, Charles W.\ Danforth
\altaffilmark{3,}\altaffilmark{6}, \\
  Margarita Rosado\altaffilmark{4}, C.-H. Rosie Chen\altaffilmark{1}}
\altaffiltext{1}{Astronomy Department, University of Illinois, 
        1002 W. Green Street, Urbana, IL 61801, USA;
        naze@astro.uiuc.edu, chu@astro.uiuc.edu, points@astro.uiuc.edu, 
        c-chen@astro.uiuc.edu}
\altaffiltext{2}{Institut d'Astrophysique et de G\'eophysique, Avenue de 
    Cointe 5, B 4000 Li\`ege, Belgium; naze@astro.ulg.ac.be}
\altaffiltext{3}{Department of Physics and Astronomy, Johns Hopkins 
    University, 3400 N. Charles Street, Baltimore, MD 21218; danforth@pha.jhu.edu}
\altaffiltext{4}{Instituto de Astronom\'{\i}a IA - UNAM, Apartado 70-264, 
   04510 Mexico D.F., Mexico; margarit@astroscu.unam.mx}
\altaffiltext{5}{Research Fellow FNRS (Belgium)}
\altaffiltext{6}{Visting astronomer, Cerro Tololo Inter-American Observatory}

\begin{abstract} 

Massive stars are expected to produce wind-blown bubbles in the 
interstellar medium; however, ring nebulae, suggesting the 
existence of bubbles, are rarely seen around main-sequence O stars.  
To search for wind-blown bubbles around main-sequence O stars, 
we have obtained high-resolution {\it Hubble Space Telescope} 
WFPC2 images and high-dispersion echelle spectra of two pristine 
\hii\ regions, N11B and N180B, in the Large Magellanic Cloud.
These \hii\ regions are ionized by OB associations that still contain
O3 stars, suggesting that the \hii\ regions are young and have 
not hosted any supernova explosions.  Our observations show that
wind-blown bubbles in these \hii\ regions can be detected 
kinematically but not morphologically because their expansion 
velocities are comparable to or only slightly higher than the 
isothermal sound velocity in the \hii\ regions.  
Bubbles are detected around concentrations of massive stars,
individual O stars, and even an evolved red supergiant (a fossil
bubble).  Comparisons between the observed bubble dynamics and
model predictions show a large discrepancy (1--2 orders of magnitude)
between the stellar wind luminosity derived from bubble
observations and models and that derived from observations
of stellar winds.  The number and distribution of bubbles in N11B 
differ from those in N180B, which can be explained by the difference 
in the richness of stellar content between these two \hii\ regions.
Most of the bubbles observed in N11B and N180B show a blister-structure,
indicating that the stars were formed on the surfaces of dense clouds.
Numerous small dust clouds, similar to Bok globules or elephant
trunks, are detected in these \hii\ regions and at least one of
them hosts on-going star formation.

\end{abstract} 

\section{Introduction} 
The fast wind of a massive star can sweep the
ambient medium into a shell and form a wind-blown bubble
(c.f. Weaver et al.  1977).  If the ambient density is sufficiently high, the
swept-up shell should appear as a ``ring nebula" in \ha\ images.  Ring
nebulae are frequently seen around evolved stars with copious stellar
winds, such as Wolf-Rayet (WR) stars and luminous blue variables
(LBVs); however, ring nebulae are rarely seen around unevolved massive
stars, such as main-sequence O stars \citep{Chu91}.  In fact, no
main-sequence O stars in the Galaxy possess well-defined optical ring
nebulae.

The paucity of optical ring nebulae around unevolved massive
stars is puzzling, as early-type O dwarfs may have stellar winds with
terminal velocities of $\sim$2,000 km~s$^{-1}$ and mass loss rates of a
few $\times$10$^{-6}$ M$_\odot$~yr$^{-1}$ \citep{Petal90,deJager88}.  
Three possible explanations exist: (1) no careful, systematic searches
for ring nebulae around main-sequence massive stars have been
conducted, (2) main-sequence massive stars preferentially reside in low
density media, and (3) the microstructure of the interstellar medium
prohibits the formation or the visibility of wind-blown bubbles
\citep[e.g.,][]{Metal84}.

To search for bubbles blown by unevolved massive
stars, we have chosen to observe two young OB associations in dense
\hii\ regions in the Large Magellanic Cloud (LMC).  OB associations are
selected, so that a large number of stars can be imaged in the same
frame.  The ``young age" criterion is used to avoid confusing supernova
remnants (SNRs) and superbubbles, which are formed by evolved stars.  
The dense \hii\ region environment is required to ensure a reasonably
dense medium around the massive stars.  The two OB associations we have
selected are LH10 and LH117 \citep{LH70} in the \hii\ regions 
N11B\footnote{N11B is a compact \hii\ region in the N11 \hii\ complex. 
While N11 contains a known SNR, N11L, it is at 180~pc away and is 
totally unrelated to N11B.} and
N180B \citep{hen}, respectively.  Both OB associations contain O3 stars
(Parker et al.\ 1992, hereafter PGMW; Massey et al.\ 1989, hereafter 
MGSD).  The presence of O3 stars indicates that even
the most massive stars have remained intact, thus suggesting a young
age and enhancing the likelihood of a real absence of previous
supernovae, and justifying the ``pristine" state of the \hii\ region.

The observations we have carried out for the \hii\
regions N11B and N180B include high-resolution images obtained with the
$Hubble$ $Space$ $Telescope$ Wide Field/Planetary Camera 2 ($HST$
WFPC2) and high-dispersion long-slit spectra obtained with an echelle
spectrograph on the 4~m telescope at Cerro Tololo Inter-American
Observatory (CTIO).  Despite the high spatial resolution of the $HST$
observations, few ring nebulae are found around main-sequence massive
stars in LH10 and LH117.

In this paper we report the analysis of
$HST$ and ground-based observations of the \hii\ regions N11B and N180B
in the LMC, and use these images and spectra to investigate the issue
of ring nebulae around main-sequence O stars.  The observations are
described in \S 2, the analysis and results are reported in \S 3, 
and the conclusions are discussed in \S 4.

\section{Observations} 
The datasets used in this study include: (1)
images taken with the {\it HST} WFPC2, and (2) high-dispersion echelle spectra
taken with CTIO 4~m
telescope.  The high-dispersion spectra are useful to diagnose
expanding shells that are not morphologically identifiable.

\subsection{{\em HST} WFPC2 Images} 
{\em HST} WFPC2 images of N11B
and N180B were taken on 1999 May 12 and 1998 April 29, respectively,
for the Cycle 6 program 6698.  For each \hii\ region, the observations
were made through the $F502N$ filter for $2\times 600$~s and the $F656N$
filter for $2\times500$~s.  The $F502N$ filter, centered at 5012.2 \AA\
with a FWHM of 26.8 \AA, includes only one nebular line, the \oiii\
$\lambda$5007 line.  The $F656N$ filter, centered at 6563.7 \AA\ with 
a FWHM of 21.4 \AA, may include the \ha\ line and the neighboring 
\nii\ lines.  At the LMC's radial velocity, $\sim$300 \kms, the 
\nii\ $\lambda$6583 line is red-shifted further away from the red 
edge of $F656N$'s bandpass, but the \nii\ $\lambda$6548 line
is red-shifted further into $F656N$'s bandpass.  
The filter transmission for the red-shifted \nii\ $\lambda$6548 is 
95\% that at the red-shifted \ha.  Using the \nii/\ha\ intensity 
ratio measured from our echelle spectra (see Section 2.2), we find 
that the \nii\ $\lambda$6548 line emission contributes to 1-2\% of the 
$F656N$ flux of N11B, and 2-4\% of that of N180B.  This \nii\ 
contamination has been corrected in our \ha\ flux measurements. 

The calibrated WFPC2 images were produced by the standard {\it HST} pipeline.
We process them further with IRAF and STSDAS routines.  The images
taken with the same filter were combined to remove cosmic rays and to
produce a total-exposure map.  The combined \oiii\ and \ha\ images were
then corrected for the intensity- and position-dependent charge
transfer efficiency by applying a linear ramp with a correction factor
chosen according to the average counts of the sky background
\citep{Ho95}.  Following the procedures
for narrowband WFPC2 photometry\footnote{available at
http://www.stsci.edu/instruments/wfpc2/Wfpc2\_faq/wfpc2\_nrw\_phot\_faq.html}, 
images were first divided by their
total exposure time, and then multiplied by the PHOTFLAM parameter of
the header, in order to get flux densities.  To obtain fluxes, we multiplied
the flux densities by the filter rectangular width calculated with SYNPHOT: 
35.8 \AA\ for the $F502N$ filter and 28.3 \AA\ for the $F656N$ filter.

Figs.~\ref{n11}a and \ref{n11}b show the large-scale environments of
N11B and N180B with the WFPC2 fields marked.  Figs.~\ref{n11}c and 
\ref{n11}d show the WFPC2 \ha\ images of N11B and N180B marked with the 
identifications and spectral types of the most relevant stars in their OB 
associations (LH~10 in N11B, PGMW; LH~117 in N180B, MGSD).
To better show the morphological features, we present the WFPC2
\ha\ images of the two nebulae individually in Figs.~\ref{n11ha} 
and \ref{n180ha}.  To show the excitation variations, the \oiii\ 
images and \oiii/\ha\ ratio map of N11B and N180B are shown in 
Figs.~\ref{n11o3} and \ref{n180o3}, respectively.

\subsection{CTIO 4 m Echelle Spectra}

Echelle spectra were obtained with the CTIO 4 m telescope in 2000 January and 
December.  The spectrograph was used with a 79 line mm$^{-1}$ echelle grating and 
the long-focus red camera.  We observed a single order centered on the 
\ha\ line by inserting a post-slit \ha\ interference filter and replacing
the cross-dispersing grating with a flat mirror.  A Tek 2048 $\times$ 2048
CCD with 24 $\mu$m pixel$^{-1}$ was used to record the images. 
 This provided a spectral sampling of 0.08 \AA\ pixel$^{-1}$ and 
a spatial sampling of 0\farcs26 pixel$^{-1}$.  
The wavelength coverage, limited by the \ha\ filter and the echelle
order, was 125 \AA; the spatial coverage, limited by the optics,
is $\sim$200$''$.  The angular resolution, determined by 
the FWHM of the seeing, was approximately 1\arcsec.  A slitwidth of
 1\farcs65 was used.  The resultant spectral resolution, measured from
 the Th-Ar lamp lines is about 13 \kms\ FWHM.

The journal of observations is given in Table~\ref{tabexp}.  When
multiple exposures existed for a given position, they were combined
taking into account their exposure times\footnote{Two observations EW of PGMW 3168
were taken. Since they were not centered at the same position, we did not 
combine them and analyze them individually}.  On the whole, nine slit
positions were observed in N11B and four in N180B.  They were all
oriented either north-south or east-west; their exact positions
were indicated by the arrows in Figs.~\ref{n11}c and \ref{n11}d.
The identifications of lines detected in the echelle observations
are shown in Fig.~\ref{spec_lines}.  The spectral lines include 
nebular \ha\ and \nii\ lines, and telluric OH lines \citep{Oetal96}.
The rest wavelengths of the nebular lines we have adopted are
\ha\ $\lambda$6562.7885 \AA, and \nii\ $\lambda$6548.0800 \AA\ and 
$\lambda$6583.4540 \AA\ \citep{Spyr95}. 

\section{Analysis and Results} 

For each nebula, we first describe the overall \ha\
morphology and surface brightness variations, and derive the rms
electron density for various regions.  We then identify morphological
features that might have been produced by stellar winds interacting
with the ambient interstellar medium, as seen in the WFPC2 \ha\ images.  
We further use the \oiii/\ha\ ratio map to determine whether a
morphological feature has an anomalous \oiii/\ha\ ratio.  An elevated
\oiii/\ha\ ratio is indicative of a higher temperature or a shock
excitation, while a lower \oiii/\ha\ ratio suggests a lower excitation
or ionization.  Next we describe the kinematic features detected in 
the echelle spectra, fit Gaussian components to the line profiles,
and use the resultant velocities to identify expanding structures and 
to determine expansion velocities.  Finally, we combine the morphological
and kinematic information and present an integrated view of the
structure of the \hii\ region.
 
\subsection{N11B} 

\subsubsection{Surface Brightness and Density} 

The \hii\ region N11B is a young star-forming region dotted with 
numerous bright-rimmed dusty features.  The OB association LH10 has a 
high concentration of stars at the southwestern part of N11B, centered 
near 4$^{\rm h}$56$^{\rm m}$44$^{\rm s}$, $-$66$^\circ$25$'$00$''$ (see
Fig.~\ref{n11}c).  To the north of this concentration is the brightest
\ha\ emission region mixed with dusty features and connected with
several long filaments.  The other members of LH10 are loosely
distributed within N11B, together with small ($\sim$1 pc in size) 
dust clouds that are similar to Bok globules \citep{Getal99} or 
elephant trunks \citep{Hetal96}.  The two most prominent small
dust clouds are situated close to luminous stars, so their surfaces 
are ionized and emit strongly in \ha: the kiwi-shaped dust cloud to 
the west of PGMW 3204 and PGMW 3209, and the Y-shaped dust cloud to the 
west of PGMW 3223 (see Fig.~\ref{n11}c).  Interestingly, the tip of the 
Y-shaped dust cloud harbors a bright star, and our echelle observations 
(\S3.1.3) show that this bright star has a strong \ha\ emission line.  
Several smaller dust clouds are present but at larger distances 
from luminous O stars, so their \ha\ surface brightnesses are not
as high as those close to luminous O stars. 

To quantitatively analyze these images, we have measured the \ha\ 
surface brightness in several regions of interest.  
We then compute the emission measure, $EM$, using:
$EM = 2.41\times10^3~T^{0.92}~S({\rm H}\alpha)$~~pc~cm$^{-6}$, where 
$T$ is the electron temperature and $S(H\alpha)$ is the intrinsic surface 
brightness in units of ergs cm$^{-2}$ s$^{-1}$ sr$^{-1}$ \citep{peim}.
Adopting an electron temperature of 10$^4$ K and assuming that the
depth of emitting material along the line of sight is similar to
the observed size of a morphological feature, we may further 
estimate the rms electron density of the emitting material.
If the feature is not spherical, the observed width (the shorter
dimension) and length (the longer dimension) are used as the 
path-length of emission to determine the upper and lower limits of 
the rms density, respectively.

We assume that the density structure of N11B consists of local
density enhancements superposed on a global, uniform component.
For the global component, we have measured the surface brightness 
of a featureless region at 9\arcsec\ east and 1\farcs4 north of 
PGMW 3168, and derived 
$S(H\alpha)$ = 9.6$\times$10$^{-4}$ ergs cm$^{-2}$ s$^{-1}$ sr$^{-1}$, 
corresponding to an EM$_{global}$ of 1.1$\times$10$^3$ cm$^{-6}$ pc,
taking into account an interstellar absorption with an E(B-V) value of 0.17 (PGMW).
The size of N11B, measured from the \ha\ image in Fig.~\ref{n11}a, 
is 257\arcsec\ $\times$ 140\arcsec, or 64 pc $\times$ 35 pc for a 
distance of 50 kpc.  The rms electron density of the global 
component in N11B is thus 13 to 18 cm$^{-3}$ in N11B.  
We have measured the density enhancements in several arcs and 
the bright ionized surface of the kiwi-shaped dust cloud.  
The intrinsic surface brightness, excess emission measure ($EM - EM_{global}$), 
size of the feature, and the range of rms density of these 
features are listed in Table~\ref{densn11}.  It can be 
seen that the moderately bright arcs and filaments have rms densities 
of 50-150 cm$^{-3}$, while the bright ionized surfaces of dust clouds 
have rms densities several times higher.

\subsubsection{Morphology and Excitation} 

Whereas there are a large number of O stars in N11B, only two of them 
(PGMW 3120 and PGMW 3160) are surrounded by filamentary features that 
are commonly identified as ring nebulae.  The star PGMW 3120, classified
as O5.5V((f*)) by PGMW, is surrounded in the northeast by a small arc-like 
structure at $\sim5''$ from the star.  The arc does not stand out in the
\oiii/\ha\ ratio map, suggesting a similar excitation to the background 
\hii\ region, or a lack of strong shock excitation.  The star PGMW 3160, 
a K I star, is in a faint void bordered in the southwest by a bright 
filament $\sim20''$ long.  This filament has a lower \oiii/\ha\ ratio 
than the background \hii\ region; however, the filament does not appear 
to be associated with an obvious dusty feature.

A few stars are surrounded by faint filaments, e.g., PGMW 3239 (B2V),
PGMW 3157 (BC1Ia), and PGMW 3102 (O7V).  Some stars, e.g., PGMW 3089
(O8V), are inside regions with suppressed surface brightness, 
indicating lower densities around the star, as in a cavity.  None of 
these features stand out in the \oiii/\ha\ ratio map;  therefore, 
from their excitation alone, they are indistinguishable from the 
background \hii\ region, indicating that they are not shock-excited.  

%The \oiii/\ha\ ratio in N11B is high near early O stars, $\sim1.0$, 
%and decreases outwards.  
While no sharp, high-\oiii/\ha\ features are detected, many
prominent low-\oiii/\ha\ features are present, and most of them are 
associated with dusty features.  The 
correlation between low \oiii/\ha\ ratios and dusty features can be 
illustrated by the analysis of an isolated dust cloud.  We use the 
kiwi-shaped dust cloud as an example.  Figure~\ref{bird}
shows that the low-\oiii/\ha\ region delineates the surface of
the dust cloud.  The \ha\ and \oiii\ surface brightness profiles
along a north-south cut across the kiwi-shaped dust cloud show that 
the \ha\ emission and \oiii\ emission on the cloud surface peak 
at the same location, but the \oiii\ emission drops off faster 
into the dust cloud than the \ha\ emission, causing a dip in the
\oiii/\ha\ ratio.  Therefore, the low \oiii/\ha\ ratios on the 
surface of a dust cloud is caused by a lower ionization and
excitation.  This kiwi-shaped dust cloud and the Y-shaped
dust cloud are both strongly ionized on the sides facing the
O3III(f*) star PGMW 3209, the earliest and most luminous O 
star in the vicinity.  These dust clouds are most likely 
illuminated and ionized by this O star.

\subsubsection{Kinematic Properties} 

To identify kinematic features caused by wind-ISM interactions,
we examine the velocity profiles of both \ha\ and \nii\ 
$\lambda$6584 lines.  The \nii\ line shows line splitting more 
easily because its thermal width is smaller than that of \ha, but 
the \nii\ line is much weaker than the \ha\ line and is usually 
noisier.  Variations of ionization condition (\nii/\ha\ ratio) 
among different velocity components also contribute to differing
appearances of the \ha\ and \nii\ line profiles.
These effects are illustrated in Fig.~\ref{speccut}, where two
pairs of \ha\ and \nii\ line profiles are shown.

We fit Gaussian components to both \ha\ and \nii\ profiles; the
resultant velocity components plotted along each slit are presented
in Fig.~\ref{specn11}.  To compare the kinematic and morphological
properties, in Fig.~\ref{split11} we mark the regions with more than 
one velocity component along the slits, with the brighter components
marked in thick solid lines and the weaker components in thick 
dashed lines.  All quoted velocities are heliocentric.  Four 
echellograms of the \nii\ line with line-splitting are shown in 
Fig.~\ref{echelle}.  

A completely different view emerged when we examined the echelle 
spectra and used the kinematic information to identify features 
caused by wind-ISM interactions.  The kinematic structure of N11B 
has been studied by \citet{Retal96} using imaging Fabry-Perot 
observations at an 18$''$ resolution (9\arcsec\ pixel$^{-1}$).  
They reported six expanding 
regions, designated a-f. To study these expanding regions at a higher 
angular resolution, we have chosen our first six slit positions 
centered on stars in these regions.  The other slit positions were 
selected to study morphologically identified features: the Y-shaped 
dust cloud and the small ring nebula around PGMW 3120.
Below we review the kinematic features along individual slits. 
The spectral types of the reference stars and the names of the
expanding regions of \cite{Retal96} are noted in parentheses.

\begin{enumerate} 

\item EW of PGMW 3204 (O6-7V; Region a): \\
The \ha\ and \nii\ lines are significantly broader in the region
near the O3III(f*) star PGMW 3209.  The \ha\ and \nii\ lines can be
fit by two Gaussian components centered at 288 \kms\ and 310 \kms.
The velocity plot shows a clear expanding shell structure, with
an expansion velocity of $\sim10$ \kms.
This expanding shell can be traced from 2\arcsec\ west to 54\arcsec\
east of PGMW 3204, corresponding to a linear size of 14 pc.  
Interestingly, the kiwi-shaped dust cloud is
located right at the western edge of the expanding shell, suggesting
again that this dust cloud is physically close to PGMW 3204.
Despite the clear expanding shell structure revealed in the velocity
plot, there are no morphological features in the WFPC2 \ha\ image that 
can be unambiguously identified as a ring nebula or an expanding shell.
Outside this expanding shell, both \ha\ and \nii\ lines can be fit
well by a single Gaussian component, centered at 298 \kms\ on the east
side and 293 \kms\ on the west side of the shell.  The western end of 
the slit reaches the bright \hii\ region north of PGMW 3120.  
The velocity profiles in the bright \hii\ region are more or less 
Gaussian, peaking at 297 \kms, but additional faint components are 
detected in the region bounded by the two arcs extending from the 
\hii\ region to the east.

\item EW of PGMW 3224 (O6III; Region b):\\
No clear line splitting is seen in the spectra along the slit, but an 
additional faint blue component sometimes creates a small asymmetry 
in the line profiles.  From 63\arcsec\ east to 7\arcsec\
west of PGMW 3224, both lines display a faint blue component at 
282 \kms, while the brighter component peaks at 303 \kms.  Further
east, the \ha\ line shows only one peak at 299 \kms.  Further west,
the line profiles peak at 299 \kms\ but become asymmetric again at 
38\arcsec\ to 60\arcsec\ west of PGMW 3224, in the vicinity of 
PGMW 3160 (early KI). At the western end of this slit position, 
the \ha\ line shows an additional component at red-shifted velocities; 
this component is too faint to be detected in the \nii\ line.

\item EW of PGMW 3168 (O7II(f); Region c):\\ 
Along this slit position, no clear line splitting is seen, but the 
peak velocities show a large scatter and the line profiles are often
asymmetric, indicating multiple velocity components.  At 133\arcsec\ 
to 50\arcsec\ east of PGMW 3168, both lines show Gaussian profiles 
centered on 298 \kms.  At 50\arcsec\ to 23\arcsec\ east of the star, 
the \ha\ line is broad and shows large velocity variation, while the 
\nii\ line profiles are noticeably asymmetric and require two velocity
components in the spectral fits.  The spectral fits to the \nii\ line
show an expanding shell with an expansion velocity of 12 \kms\ and 
a diameter of $\sim40''$ (or 10 pc).  This shell might be associated 
with the uncataloged star between PGMW 3223 and PGMW 3224.
Near PGMW 3168 and to its west, the \nii\ line also presents a main 
component (at 298 \kms) and a fainter blue component (at 283 \kms), 
although the \ha\ line shows Gaussian profiles centered at 295 \kms. 
Further west, near the bright \hii\ region, the \ha\ line also show
asymmetric profiles indicating velocity components at 290 and 307 \kms.
Still further west (more than 67\arcsec\ west of the star), both lines 
seem split but the red component is quite faint.  This kinematic
feature does not have any morphological counterparts in the WFPC2 images.

\item NS of PGMW 3058 (O3V((f*)); Region d)\footnote{Note that Rosado 
et al.'s (1996) 
Region d is centered on PGMW 3061, about 22$''$ south of PGMW 3058.}:\\ 
From 8\arcsec\ north of PGMW 3058 northwards, both \ha\ and \nii\ lines 
exhibit clear line-splitting.  It is interesting to note that a bright 
knot appears in the red component at 25$''$ north of PGMW 3058,
where the WFPC2 image shows a bright filament.  (An example of
such emission knot can be seen in the echelle image of PGMW 3053
slit in Fig.~\ref{echelle}.)
South of the star, the \ha\ line is clearly split with extreme 
velocities reaching 283 and 325 \kms, indicating an expansion 
velocity of $\sim$20 \kms.  At 13\arcsec\ to 40\arcsec\ south of 
the star, in the vicinity of PGMW 3061 (O3III(f*)), an additional, 
high-velocity component at 357 \kms\ is detected.  It is not clear 
whether this high-velocity feature is accelerated by the powerful 
stellar wind of PGMW 3061 because similar high-velocity feature is 
detected along the NS slit at PGMW 3120 only at a declination 20$''$ 
south of PGMW 3061 (see description of the slit NS of PGMW 3120 below).

\item NS of PGMW 3053 (O5.5I-III(f); Region e):\\ 
This slit position is only 8$''$ east of the slit along NS of 
PGMW 3058, thus shows a very similar kinematic structure.
The line-splittings and high-velocity features detected
along the slit centered on PGMW 3058 are all confirmed in
the spectra along the slit centered on PGMW 3053.  The bright
emission knot in the red component that corresponds to a bright
filament to the north of PGMW 3053 is clearly seen in Fig.~\ref{echelle}.

\item NS of PGMW 3157 (BC1Ia; Region f):\\ 
To the north of the star, the line profiles are mostly Gaussian and 
centered at 297 \kms; however, at 43\arcsec\ to 56\arcsec\ north of the 
star, an additional faint blue component appears at 284 \kms\ and the main
component becomes slightly red-shifted to 301 \kms.  This region,
as marked in Fig.~\ref{split11}, corresponds to the arc feature around
PGMW 3160 (early KI).  Similar velocity structure is seen in the vicinity 
of PGMW 3160 in the echelle slit EW of PGMW 3224.
From PGMW 3157 southward to 29\arcsec\ south, an additional red 
component is present at $\sim310$ \kms.  This is the region to the south
of a bright filament (see Fig.~\ref{split11}).  Similar velocity structure
in the vicinity is seen in the observation along the slit EW of PGMW 3204.

\item EW of PGMW 3223 (O8.5IV):\\ 
This slit position is selected to sample the Y-shaped dust cloud at 
$\sim8''$ west of PGMW 3223.  The line profiles are clearly split into 
287 and 306\kms\ over an extent of $\sim52''$ (or 13 pc) around 
PGMW 3223, with the Y-shaped dust cloud situated on the western edge.
The echelle spectrum of the Y-shaped dust cloud also shows an embedded
star with bright \ha\ emission.  This \ha-emission star, marked as 
``\ha\ star'' in Figs.~\ref{n11}c and ~\ref{specn11}, is probably 
a newly formed massive star.
Similar velocities are also recorded at 67 to 86\arcsec\ west of PGMW 3223, 
corresponding to extensions of the bright \hii\ region.  Everywhere else, 
the line profiles are purely Gaussian and are centered at $\sim$296 \kms.

\item EW and NS of PGMW 3120 (O5.5V((f*))):\\
These two slit positions are selected to study the ring nebula around
PGMW 3120.  As seen in the WFPC2 \ha\ image (Figs.~\ref{n11}c and 
\ref{n11ha}), this ring nebula is incomplete with a radius of 
3$''$--6$''$.  Both the \ha\ and \nii\ lines are clearly split,
with the two components at 281 and 306 \kms.
However, the spatial extent of the region with line-splitting is from
12\arcsec\ east to 26\arcsec\ west of the star and from 7\arcsec\ north
to at least 40$''$ south of the star.  This expanding shell is much 
more extended than the morphologically identified ring nebula, and
encompasses the nearby O7V star PGMW 3102 (see Figs.~\ref{specn11} 
and \ref{echelle}).  The small \ha\ arc structure is probably fortuitous.
At 40\arcsec\ to 54\arcsec\ south of PGMW 3120, high-velocity gas at
$\sim$347 \kms\ is seen.  This region is within Rosado et al.'s (1996) 
region d, and the velocity features are similar to those within region d 
sampled by the NS echelle slits centered on PGMW 3058 and PGMW 3053.
The eastern part of the EW slit samples part of Rosado et al.'s
(1996) region a.  Near stars PGMW 3204 and 3209, the \ha\ line is
split into 287 and 308 \kms, similar to those seen in the observation
along the EW slit centered on PGMW 3204.

 \end{enumerate}

\subsubsection{Integrated View of N11B}

The analysis of the echelle observations of N11B shows clearly that
expanding shells and high-velocity, accelerated material exist in 
this young \hii\ region.  Most of the expanding shells do not
show identifiable morphological counterparts, while the morphologically
identified ring nebulae do not show expanding shell structures.

Most of the expanding shells encompass
groups of massive stars; only a few surround single massive stars.
Some irregular expanding features with expansion velocities 10--15 
\kms\ are detected, but no clear association with massive stars can be 
identified.  The expanding shells and high-velocity features associated
with recognizable massive stars are described below and summarized in 
Table~\ref{bubsum}.  The spectral types and expected wind luminosity,
calculated using the wind velocities from \citet{Petal90} and the
mass loss rates from \citet{deJager88}, are also given in this table.

The shell encompassing the large concentration of massive stars
at the SW corner of N11B corresponds to Rosado et al.'s region d.
This shell has the largest expansion velocity, $\sim$20 \kms, and 
the largest dimension, $\ge$20 pc (see the velocity plots of NS of 
PGMW 3053, PGMW 3058, and PGMW 3120 in Fig.~\ref{specn11}).  
The south boundary of this shell is unknown, as it is not covered in 
our echelle slits.  
The second largest shell corresponds to Rosado et al.'s region a.
This shell encompasses the concentration of stars including PGMW 
3204, PGMW 3209, and PGMW 3223.  Its expansion velocity is $\sim$10 
\kms\ and its diameter is $\sim$14 pc (see, e.g., the velocity plot 
of EW of PGMW 3204 in Fig.~\ref{specn11}).  

One expanding shell might be associated with a single star,
the uncataloged bright star between PGMW 3223 and PGMW 3224.
This shell has an expansion velocity of $\sim$12 \kms, and
a diameter of $\sim$10 pc (see the velocity plot of PGMW 3168
in Fig.~\ref{specn11}).  
A high-velocity feature, red-shifted
by $\sim$60 \kms\ and extending over $\sim$7 pc, is detected near
PGMW 3061, an O3III(f*) star (see the velocity plots of NS of 
PGMW 3053 and PGMW 3058 in Fig.~\ref{specn11}).  It is possible that 
PGMW 3061 is responsible for its acceleration.  
The most interesting expanding structure around a single massive star
belongs to PGMW 3160, an early KI supergiant.  An expanding blister
structure, with a size of $\sim$5 pc and expansion velocity of $\sim15$
\kms, is detected in the vicinity of this red supergiant (see the 
velocity plot of NS of PGMW 3157).  This expanding structure is most 
likely produced by the O star progenitor of PGMW 3160 during the main 
sequence stage.

Three characteristics appear to be common among the expanding shells 
in N11B.  First, shells encompassing small numbers of massive 
stars have sizes of 10--15 pc and expansion velocities of 10--15 \kms, 
barely supersonic for an isothermal sound velocity of 10 \kms\ at 10$^4$ K.
Only the shell encompassing a large number of stars is larger and
expands faster.  Second, Most of the expanding shells have the massive 
stars displaced from the shell centers, indicating a blister-like structure.
Finally, some shells might be merging together and the boundaries between 
them become unclear.

\subsection{N180B}

\subsubsection{Surface Brightness and Density}

The most remarkable feature in the WFPC2 \ha\ image of N180B is 
the prevalence of dust clouds in different sizes and shapes.  
There are 30-pc-long dust streamers running across the face of
the \hii\ region, prominent antenna-shaped dust cloud at the 
northeastern corner of the field, large patches of dust clouds at 
the northwestern and southwestern corners of the field, and 
Bok-globule-like small dust clouds embedded in ionized gas near 
the large dust clouds.  These dusty features amply portray a 
young star formation region.

The \ha\ surface brightness is heavily modulated by the large-scale
dusty features.  The brightest emission region in N180B is located 
in the southwestern quadrant of N180B.  Its V-shaped rim is dotted
with four bright O and B stars: MGSD 152, MGSD 184, MGSD 187, and 
MGSD 197 (see Fig.~\ref{n11}d for the positions and spectral types).  
As shown below, this higher surface brightness is caused by a higher 
density in this region.

Again, we have adopted a simplistic two-density-component model for 
the structure of N180B: a global uniform component and local density 
enhancements.  We measured the surface brightness of a featureless 
region at 3\farcs5 north and 17\arcsec\ west of the star MGSD 119
to represent the global component.  Adopting a diameter of 225\arcsec\
(or 56 pc) for N180B and an E(B-V) value of 0.12 (MGSD), 
we derive a rms density of $\sim$9.5 cm$^{-3}$.
We have measured the surface brightness of the bright, V-shaped 
emission region and the arc around MGSD 218.  The rms densities of 
these two regions are a few $\times$10 cm$^{-3}$.  The results are
detailed in Table~\ref{densn11}.

\subsubsection{Morphology and Excitation} 

Of all stars in the WFPC2 \ha\ image of N180B, only two stars,
MGSD 174 and MGSD 218, are surrounded by small (5$''-10''$) arcs 
or circular filaments that can be identified as ring nebulae.  
No spectral classification was available for these two stars,
but the UBV photometric data and extinction reported
by MGSD suggest that the spectral type of MGSD 174 is B2V and 
MGSD 218 B5V.  A few stars may 
be associated with faint nebular features, such as a small 
circular emission region around MGSD 82 and the ``comet-tail'' 
of MGSD 106; however, these morphologies do not suggest a bubble 
structure.

The average \oiii/\ha\ ratio of N180B is high, $\sim$0.9, comparable
to that of N11B.  In the \oiii/\ha\ ratio map of N180B, most dust clouds, 
but not the long dusty streamers, are seen as 
regions with low \oiii/\ha\ ratio, $\sim$0.4.  The dusty streamers 
across the face of the \hii\ region show \oiii/\ha\ ratios similar 
to that of the overall \hii\ region itself.  The \oiii/\ha\ ratio 
map of N180B appears much more uniform than that of N11B.  Again,
no sharp, shock-excited features are identified.

The small dust cloud near MGSD 168 has \oiii/\ha\ ratios of 
0.6--0.9, which are higher than those in N11B, such as the kiwi-shaped
dust cloud.  This difference is probably caused by the contamination
of background/foreground emission.  MGSD 168 is a mid B star, so the
\ha\ emission from its neighboring dust cloud is weak compared to the 
background \ha\ emission from gas ionized by early O stars, while
the kiwi-shaped dust cloud in N11B is ionized by early O stars
and its emission is much brighter than the background emission.
Therefore the low \oiii/\ha\ ratio in the ionization front can 
be more easily seen in dust clouds near early O stars than later 
O or B stars.

The arc around MGSD 218 has a low \oiii/\ha\ ratio, 0.45, indicating
a low excitation at an ionization front.  This kind of arc nebulae 
is frequently seen around early B main sequence stars near dust clouds,
e.g., LSS~3027 in the Galaxy \citep{chu83}.  MGSD 218 is a mid B main 
sequence star and could be luminous enough to produce a small \hii\ region.
The filamentary nebula around MGSD 174 may also have a low \oiii/\ha\ 
ratio, but the S/N ratio of
this region (imaged by the Planetary Camera of WFPC2) is too low to
confirm the low \oiii/\ha\ ratio.

Many nebular morphological features ``disappear'' into the background
in the \oiii/\ha\ ratio map.  The small ``comet-tail'' nebulosity 
of MGSD 106 and the large (20 pc in size), bright, V-shaped emission 
region both show \oiii/\ha\ ratios similar to the average ratio of
the entire \hii\ region.   It is interesting to note that MGSD 118
(O4 IIIf* star) is surrounded by a small emission region (size = 
3\arcsec) with a very high \oiii/\ha\ ratio ($\sim$1.5), but this
small nebula does not show any peculiar kinematic properties 
(see Section 3.2.3).

\subsubsection{Kinematic Properties}

The kinematic structure of N180B is much more quiescent than that of
N11B.  Line splitting is seen only in the \nii\ line within one 
expanding shell centered on MGSD 214 (O3-4).  The \ha\ line is narrow 
with an observed FWHM of $\sim$30 \kms.  Quadratically subtracting 
the instrumental FWHM of 13 \kms\ and a thermal FWHM of 21 \kms\ 
(assuming 10$^4$ K) from the observed width, we obtain a turbulent 
FWHM of only 17 \kms.

Figure~\ref{specn180} shows the velocity components of the \ha\ 
and \nii\ lines along each slit.  Regions with more than one
velocity component are marked on the \ha\ image in Fig.~\ref{split180}.
The kinematic features along each slit position are described below.

\begin{enumerate} 

\item NS of MGSD 118 (O4IIIf*):\\ 
The northern part of this slit, including the location of MGSD 118,
show narrow \ha\ and \nii\ lines centered at 245--250 \kms.  
From 28\arcsec\ south of MGSD 118 southward, the \ha\ line shows an 
additional faint blue component, which has an extreme velocity at
220--225 \kms.  This suggests an expanding ``blister", with the far
side roughly at the \hii\ region's systemic velocity and the near 
side expanding toward us with an expansion velocity of $\sim$20 \kms.
The northern edge of the region appears to be bordered by a broad 
arc-like structure.

\item EW of MGSD 140 (O3-4(f*)):\\
Along the entire slit, the \ha\ line appears to be a perfect Gaussian 
profile centered at $\sim$240 \kms, while the \nii\ line shows asymmetric
line profiles indicating the existence of at least two velocity 
components.  Spectral fits of the \nii\ line yield components separated 
by $\le$20 \kms.  The \nii\ line is weak and noisy, and thus the velocity
components show a large velocity scatter.

\item EW and NS of MGSD 214 (O3-4):\\ 
From 50\arcsec\ east to 40\arcsec\ west of MGSD 214, both \ha\ and \nii\ 
lines show a bright component at 240--245 \kms, near the \hii\ region's 
systemic velocity, and a faint blue component at 220--225 \kms.  This 
velocity pattern suggests a ``blister" expanding toward us.  The N-S
slit position shows that this expanding blister extends from $\sim$65\arcsec\
north to $\sim25$\arcsec\ south of MGSD 214.  Outside this expanding blister
region, the \ha\ and \nii\ lines are both narrow.  Velocity gradients are 
seen along both slits, from one side of the expanding blister to the other
side.  It is not clear whether the velocity variation of the bright
component is attributed largely to the velocity gradient in the \hii\ region 
or an expansion velocity variation in the receding side of the blister.

\end{enumerate}

\subsubsection{Integrated View of N180B}

N180B is much more quiescent than N11B.  Only one expanding blister
is unambiguously detected around the O3-4 star MGSD 214.  This blister
is $\sim$22 pc in diameter and expanding away from the dense cloud at 
$\sim$20 \kms\ (see Table~\ref{bubsum}).  No morphological counterpart 
of this expanding blister
can be confidently identified.  Two other stars have similar spectral 
types, MGSD 118 of type O4 IIIf* and MGSD 140 of type O3-4 (f*), but
do not show unambiguous expanding shells or blisters around them.
The FWHM of the \ha\ line in the vicinity of these two stars 
suggests a turbulent component of $\le$17
\kms\ FWHM, while the \nii\ line shows additional faint blue components.  
It is possible that these two massive stars have blisters expanding at 
$\le$10 \kms.

The overall kinematic structure of N180B exhibits two characteristics.
First, the \hii\ region itself has a velocity gradient or variation
at a level of $\sim$10 \kms\ across the nebula.  Second, the expanding
shells have a ``blister" structure expanding toward us, with the 
back side being the dense \hii\ region itself.

\section{Discussion and Conclusions} 

Massive stars are expected to produce wind-blown bubbles in the 
interstellar medium \citep{cas,Wetal77}; however, main-sequence O
stars are rarely seen surrounded by  ring nebulae 
\citep{Chu91}.  To search for wind-blown bubbles around
main-sequence O stars, we have obtained high-resolution {\it HST} 
WFPC2 images and high-dispersion echelle spectra of two pristine 
\hii\ regions, N11B and N180B, ionized by OB associations.  The
presence of O3 stars in these OB associations suggests that the 
\hii\ regions are young and have not hosted any supernova explosion.

Analyzing the nebular morphologies and kinematics, we find that 
main-sequence O stars do blow bubbles in \hii\ regions, and that
the bubbles can be detected kinematically\footnote{Note that 
the photo-evaporation of dense knots will produce line broadening 
(see simulations by Brandner et al. 2000), not the expanding shells 
we observed.} but not morphologically.
Most of these main-sequence bubbles have expansion velocities 
of 10--15 \kms, which is comparable to or slightly larger than the
isothermal sound velocity of ionized gas at 10$^4$ K.  The 
expanding bubbles generate only weak interstellar shocks as they 
expand into the ambient interstellar medium; therefore, no strong
compression of the interstellar medium is expected in these bubbles.
Lacking sharp filaments in \ha\ images, the slowly-expanding bubbles
blown by main-sequence stars cannot be identified morphologically.
Ironically, in N11B and N180B the morphological features that are 
conventionally identified as ring nebulae, such as a bright nebular
arc around a star, are not expanding shells (e.g., MGSD 218 in N180B or
PGMW 3120 in N11B).

Three types of wind-blown bubbles are seen in those young \hii\ regions:
(1) bubbles around concentrations of massive stars within which the
projected separation between neighboring stars are typically $\le$ 1 pc,
(2) bubbles around individual O stars if the distance to neighboring
stars is large enough to maintain the individuality of the bubble, and 
(3) a fossil bubble around an evolved red supergiant that is isolated.
It is conceivable that the formation of wind-blown bubbles or
superbubbles in an \hii\ region depends crucially on the spatial
distribution of the stars.  Tight concentrations of massive stars
are conducive to the formation of superbubbles even before any 
supernova explosion.  Wide separations between massive stars are
conducive to the formation of single-star, main-sequence bubbles, 
and a bubble may remain distinct even after the star has evolved 
into a red supergiant.

The observed expansion dynamics of the interstellar bubbles can
be compared to those expected in bubble models of \citet{cas} and 
\citet{Wetal77}. With the observed radius ($R$ in units of pc) and 
expansion velocity ($V$ in units of \kms), we can compute the dynamical
timescale ($t_6$ in units of 10$^6$ yr) using $t_6 = 0.6 R / V$, and further 
determine the ratio of stellar wind luminosity ($L_{36}$ in units
of 10$^{36}$ ergs~s$^{-1}$) to ambient density ($n_0$ in units
of H-atom~cm$^{-3}$) using $L_{36}/n_0 = (R/27)^5 t_6^{-3}$.
We adopt the rms densities we have determined for N11B and N180B,
15 H-atom~cm$^{-3}$ and 9.5 H-atom~cm$^{-3}$, respectively.
We have listed in Table~\ref{bubsum} the observed diameter and 
expansion velocity of the expanding shells, together with the 
derived dynamical timescale and wind luminosity.   We have also 
listed the early-type stars encompassed within each shell and 
the stellar wind luminosity expected from these stars. 
 
We will focus on the two best-defined shells in our sample.
First we consider the bubble blown by the O3--4 star MGSD 214 in N180B.
With a radius of 11 pc and an expansion velocity of $\sim$20 \kms, the
bubble around MGSD 214 has a dynamical timescale of 3.3$\times$10$^5$ yr.
Adopting N180B's rms density of $\sim$9.5 cm$^{-3}$, we find a wind
luminosity of $\sim$3$\times$10$^{36}$ ergs~s$^{-1}$ for MGSD 214.
This is a factor of 10 lower than that expected from an O3--4 star.  
Next we consider the bubble blown by PGMW 3204, 3209, and 3223
in N11B.  The star PGMW 3209, while dominated by an O3III(f*) star,
is in fact composed of at least 5 additional late O stars 
\citep{wal}.  For a radius of 7 pc and an expansion velocity of 10 \kms, 
the dynamical timescale of this bubble is 4.2$\times$10$^5$ yr.  Using a rms 
density of 15 cm$^{-3}$ for N11B, we expect a wind luminosity of 
$\sim$2.5$\times$10$^{35}$ ergs~s$^{-1}$ for PGMW 3209, two orders of 
magnitude lower than that expected from an O3III star.  This discrepancy 
is even larger, if we consider the additional stars in PGMW 3209 and
the neighboring stars PGMW 3204 and 3223.

The large discrepancy between stellar wind luminosities derived 
from observations and models of bubbles and those expected from 
the actual stellar content is very interesting.  Similar discrepancy
has been reported for circumstellar bubbles 
\citep[e.g., NGC~6888;][]{Getal96} and superbubbles \citep{Oey96}.
Our results represent the first report for small interstellar
bubbles.  This discrepancy implies that either the observed stellar
wind luminosities have been over-estimated or Weaver et al.'s 
model cannot describe the observed bubbles.  If the observed stellar 
wind luminosity is correct, our observed bubbles will be too small
and expand too slowly.  To reduce the discrepancy, we need to
increase the ambient density, but it is unlikely that we have
under-estimated the rms density of the \hii\ regions by factors of
10--100. 

One further discrepancy underscores the deficiency of bubble models. 
The dynamical timescales of the bubbles in N11B, 0.1--0.5 Myr, are 
much shorter than the age of the OB association LH10, 2--3 Myr 
\citep{wp}. A similar discrepancy is also observed in superbubbles
\citep{Oey96}. These problems need to be seriously considered in 
the future.

The content and distribution of wind-blown bubbles in N11B are different
from those in N180B: N11B has a larger number of wind-blown bubbles and 
faster-expanding bubbles.  This difference is caused by the richness of
massive stars in the encompassed OB associations.  The OB association
LH10 in N11B has 20 stars with stellar masses $M \ge 15$ M$_\odot$
(PGMW), while the OB association LH118 in N180B only has six
(MGSD).  Futhermore, LH10 has tighter concentrations of massive
stars than LH118.  Bubbles blown by concentrations of massive stars 
expand faster because of more powerful stellar winds.  Therefore, N11B 
has more bubbles and the fastest expanding bubble is around its highest 
concentration of massive stars.

One property in common among the bubbles in N11B and N180B is the
asymmetric structure -- most of the bubbles are ``blisters".  
A face-on blister can be diagnosed from the asymmetry in line profiles, 
which show a bright component near the \hii\ region's systemic
velocity and a faint component blue-shifted or red-shifted to higher 
velocities.  A blister viewed from the side can be diagnosed by the 
displacement of the responsible star from the projected bubble center.  
As the stars in N11B and N180B are still young and must be located 
close to their birth places, their blister-like bubbles suggest that
the stars are located near the surface of a dense cloud, where a 
density gradient is present; as opposed to embedded in the deep 
interior of a cloud, where the ambient density might be more uniform.

Finally, we note that numerous small dust clouds like Bok globules 
and elephant trunks are detected in N11B and N180B.  If the ambient 
interstellar gas pressure is high enough, star formation might be 
triggered in these small dust clouds.  At least one such cloud in 
N11B, the Y-shaped dust cloud near PGMW 3223, shows signs of on-going 
star formation at its tip.  An infrared survey of N11B and N180B 
would reveal whether star formation has started in other small dust
clouds.

\acknowledgments 

We would like to thank Michael Dopita for prompt review and constructive 
suggestions. This research is supported by the STScI grant GO-06698.01-95A.  
Y.N. acknowledges the support from contract P4/05 ``P\^ole
d'Attraction Interuniversitaire'' (SSTC-Belgium) and from the PRODEX
XMM-OM and Integral Projects.  CD acknowledges the travel support from CTIO.

\clearpage

%\begin{figure} 
%\epsscale{0.48} 
%\plotone{naze.fig1a.eps} 
%\plotone{naze.fig1b.eps}
%\plotone{naze.fig1c.eps}
%\plotone{naze.fig1d.eps}
%\caption{(a) Curtis Schmidt \ha\ CCD image of N11B with the WFPC2 field outlined.
%(b) Same as (a) for N180B.
%(c) {\it HST} WFPC2 \ha\ image of N11B.  Stars cataloged by PGMW are marked
%with their catalog number and spectral classification.  Slit positions 
%of the echelle spectra are indicated by arrows.
%(d) Same as (c) for N180B. The star numbers and spectral types are taken 
%from MGSD.  Images presented in this article, except for Fig.~\ref{bird}, 
%are negative images. \label{n11}} 
%\end{figure}
%\clearpage

\begin{figure} 
%\epsscale{0.48} 
%\plotone{naze.fig1a.eps}
\caption{(a) Curtis Schmidt \ha\ CCD image of N11B with the WFPC2 field outlined.
\label{n11}} 
\end{figure}
%\clearpage
\setcounter{figure}{0}
\begin{figure} 
%\epsscale{0.48} 
%\plotone{naze.fig1b.eps}
\caption{(b) Same as (a) for N180B.} 
\end{figure}
%\clearpage
\setcounter{figure}{0}
\begin{figure} 
%\plotone{naze.fig1c.eps}
\caption{(c) {\it HST} WFPC2 \ha\ image of N11B.  Stars cataloged by PGMW are marked
with their catalog number and spectral classification.  Slit positions 
of the echelle spectra are indicated by arrows.}
\end{figure}
%\clearpage
\setcounter{figure}{0}
\begin{figure} 
%\plotone{naze.fig1d.eps}
\caption{(d) Same as (c) for N180B. The star numbers and spectral types are taken 
from MGSD.  Images presented in this article, except for Fig.~\ref{bird}, 
are negative images. } 
\end{figure}
%\clearpage

\begin{figure} 
\epsscale{1.0} 
%\plotone{naze.fig2.eps}
\caption{{\it HST} WFPC2 \ha\ image of N11B.  The bar below the image
shows the greyscale for surface brightness in units of erg cm$^{-2}$s$^{-1}$.  
\label{n11ha}} 
\end{figure}
%\clearpage

\begin{figure} 
\epsscale{1.0} 
%\plotone{naze.fig3.eps} 
\caption{Same as Fig.~\ref{n11ha}, for N180B.
\label{n180ha}} 
\end{figure} 

%\clearpage
\begin{figure}
\epsscale{0.5}
%\plotone{naze.fig4.eps}
\caption{{\it HST} WFPC2 \oiii\ image (left) and \oiii/\ha\ ratio map 
(right) of N11B.  The bars below the images show the greyscales for 
surface brightness (in units of erg cm$^{-2}$s$^{-1}$) and line ratio,
respectively.
\label{n11o3}}
\end{figure}
%\clearpage 

\begin{figure}
\epsscale{0.5}
%\plotone{naze.fig5.eps}
\caption{Same as Fig.~\ref{n11o3}, for N180B.
\label{n180o3}}
\end{figure}

%\clearpage 
\begin{figure}
%\plotone{naze.fig6.eps} 
\caption{Identification of the telluric and nebular lines present 
in the echelle spectra of N11B and N180B.  The telluric lines are narrow 
and unresolved. The intensity scale is arbitrary.
\label{spec_lines}} 
\end{figure} 
%\clearpage

\begin{figure} 
\epsscale{1.0} 
%\plotone{naze.fig7.eps} 
\caption{(a) {\it HST} WFPC2 \ha\, (b) \oiii, and (c) \oiii/\ha\ ratio 
map of the kiwi-shaped dust cloud in N11B.  The size of each image is 
9\arcsec$\times$6\arcsec. 
Note that these images are presented with emission in white, unlike the
other figures. (d) Profiles of \ha\ and \oiii\ surface brightness, and the
\oiii/\ha\ ratio (multiplied by 10$^{-16}$) along a cut 
bounded by the white lines on the \ha\ image in (a).  The X-axis is in 
units of WFC pixel with a size of 0\farcs1. The profiles plot extends
further than the images shown in (a)-(c); the dashed line above the
profiles marks the 6\arcsec\ extent. \label{bird}} 
\end{figure} 
%\clearpage

\begin{figure}
%\plotone{naze.fig8.eps}
\caption{Examples of \ha\ and \nii\ line profiles.  (a) \ha\ and
(b) \nii\ profiles at 47\farcs5 south of PGMW 3053. (c) \ha\ and
(d) \nii\ profiles around PGMW 3120.
\label{speccut}}
\end{figure}
%\clearpage

\begin{figure}
%\plotone{naze.fig9.eps}
\caption{Radial velocity-position plots of the echelle slit 
positions in N11B for both \ha\ and \nii\ lines.  The radial 
velocities are heliocentric.  The \ha\ velocity components
are plotted in filled symbols, and the \nii\ in open ones.
The position axis is defined with the central star (labeled
with its PGMW number at the upper left corner) at the origin 
(0\arcsec) and increasing towards south or west.  When another
star is present in the spectrum, its position is marked by
a short vertical line labeled with its PGMW number.  Stars close
to the slit position are also marked but in parentheses.
The dashed line in each plot indicates the range covered by
the WFPC2 images.  The two observations EW of PGMW 3168
were not centered at the same position; they were analyzed 
individually and plotted in different symbols, squares and
triangles.
\label{specn11}}
\end{figure}
%\clearpage 

\begin{figure}
%\plotone{naze.fig10.eps} 
\caption{{\it HST} WFPC2 \ha\ image of N11B superimposed by spectral
features: a thin line for a Gaussian profile, thick double lines for
lines with clear splitting, and thick solid-dashed lines for
asymmetric profiles indicating additional faint components.  
The relevant reference stars are also labeled.  Note that the
widths of the lines do not represent the slitwidth, which is 
1\farcs65.
\label{split11}} 
\end{figure} 
%\clearpage

\begin{figure} 
\epsscale{0.8}
%\plotone{naze.fig11.ps} 
\caption{Four examples of echellograms of the \nii\ line.
The slit orientation and stars along the slit are marked.
Each panel is 185\arcsec\ along the horizontal direction 
and 200\kms\ along the vertical direction.
The beginning of the splitting region seen to the north
of PGMW 3053 is clearly marked by a bright knot in the spectrum.  
The line splitting around PGMW 3120 is clearly visible,
while the ``line splitting" region to the east of PGMW 3204 
shows only a broad line.  \label{echelle}} 
\end{figure} 
%\clearpage

\begin{figure} 
%\plotone{naze.fig12.eps} 
\caption{Same as Fig.~\ref{specn11}, but for N180B. The star
numbers are from MGSD.  
\label{specn180}} 
\end{figure} 
%\clearpage

\begin{figure}
%\plotone{naze.fig13.eps}
\caption{ Same as Fig.~\ref{split11}, but for N180B.  The star
numbers are from MGSD.  \label{split180}}
\end{figure}
\clearpage

\begin{table} 
\begin{center} 
\caption{Journal of Echelle Observations.\label{tabexp}} 
\begin{tabular}{lccc} 
\tableline
\tableline 
Central & Slit& Exp. & Obs. \\
Star & Orient.& Time & Date \\
\tableline
N11B&&&\\
PGMW 3204 & EW& 600s& 2000 Jan 23\\
PGMW 3224 & EW& 600s& 2000 Jan 23\\
PGMW 3168 & EW& 2$\times$600s\tablenotemark{a}& 2000 Jan 23\\
PGMW 3058 & NS& 2$\times$300s& 2000 Jan 22\\
PGMW 3053 & NS& 2$\times$300s& 2000 Jan 22\\
PGMW 3157 & NS& 2$\times$300s+10s& 2000 Jan 22\\
PGMW 3223 & EW& 600s& 2000 Jan 23\\
PGMW 3120 & NS& 2$\times$300s& 2000 Jan 22\\
PGMW 3120 & EW& 600s& 2000 Jan 23\\
\tableline
N180B&&&\\
MGSD 140 & EW& 600s& 2000 Jan 23\\
MGSD 118 & NS& 600s& 2000 Dec 10\\
MGSD 214 & NS& 600s& 2000 Dec 10\\
MGSD 214 & EW& 600s& 2000 Jan 23\\
\tableline
\end{tabular}
\tablenotetext{a}{Two observations of PGMW 3168
were taken, but they were not centered at the same position.}
\end{center}
\end{table}
\clearpage

\begin{table} 
\begin{center} 
\caption{Densities of Selected Regions in N11B and N180B. 
\label{densn11}} 
\begin{tabular}{lcccc}
\tableline
\tableline 
Region\tablenotemark{a} & S(\ha) & EM-EM$_{global}$ & Size  &rms Density\\ 
&(10$^{-3}$erg
cm$^{-2}$s$^{-1}$sr$^{-1}$)&(10$^4$cm$^{-6}$pc)&(\arcsec$\times$\arcsec)
&(cm$^{-3}$)\\ 
\tableline 
N11B &&&&\\
Near PGMW 3168(background) &0.955$\pm$0.016&(1.101$\pm$0.019)\tablenotemark{b} 
&257$\times$140&13-18\\ 
Arc S. of PGMW 3160 & 1.823$\pm$0.020 & 1.001$\pm$ 0.030& 17.8$\times$1.5
&50-160\\ 
Arc W. of PGMW 3148 & 3.280$\pm$0.028 & 2.683$\pm$ 0.037&
$11.5\times$2.2 &100-220\\ 
Arc N. of PGMW 3053 & 2.288$\pm$0.025 & 1.538$\pm$ 0.035& 7.8$\times$2.8 &90-150\\ 
Bright region of the kiwi
cloud & 5.096$\pm$ 0.127& 4.776$\pm$0.147 & 2.6$\times$0.8 &270-500\\
\tableline
N180B&&&&\\
Near MGSD 119 (background)& 0.424$\pm$0.003 & (0.489$\pm$0.004)\tablenotemark{b} 
& 225$\times$225 &9.5\\ 
S. of MGSD 214& 0.297$\pm$0.002& (0.343$\pm$0.002)\tablenotemark{c} & 
225$\times$225 &8\\ 
Arc N. of MGSD 218 & 0.696$\pm$0.007 & 0.314$\pm$0.009 & 5.4$\times$1.3 &50-100\\ 
V-shape zone between& 0.764$\pm$0.002 & 0.392$\pm$0.005 & 37.0$\times$7.0
&20-50\\ 
MGSD 187 and MGSD 197&&&&\\ 
\tableline 
\end{tabular} 
\tablenotetext{a}{The different regions can be identified in Figs,~\ref{n11}c 
and~\ref{n11}d.}
\tablenotetext{b}{This is the background reference, EM$_{global}$, and the size 
of this feature is the nebula's.}
\tablenotetext{c}{This EM is not background subtracted.}
\end{center} 
\end{table} 
\clearpage

\begin{table} 
\begin{center} 
\caption{Summary of Expanding Shells Properties.\label{bubsum}} 
\begin{tabular}{llccccc} 
\tableline
\tableline 
Ident.&Stellar content &L$_w$(stars)&Diam.& V$_{exp}$&t$_6$&L$_w$(model)\\
 & &(10$^{36}$~erg~s$^{-1}$)&(pc)&(\kms)&(10$^6$~yr)&(10$^{36}$~erg~s$^{-1}$)\\
\tableline
N11B-1&all stars at the SW of N11B&&$\ge$20\tablenotemark{a}&20&&\\
N11B-2&PGMW 3204 (O6-7V)&1.03&14&10&0.42&0.24\\
&PGMW 3209A (O3III(f*))\tablenotemark{b}&15.8&&&&\\
&PGMW 3209B (O9V)&0.14&&&&\\
&PGMW 3209C (O7V)&0.59&&&&\\
&PGMW 3209D (O9V)&0.14&&&&\\
&PGMW 3209E (O9.5V)&0.03&&&&\\
&PGMW 3209F (O9.5V)&0.03&&&&\\
&PGMW 3223 (O8.5V)&0.08&&&&\\
N11B-3&uncataloged star between&&10&12&0.25&0.21\\
& PGMW 3223 and 3224&&&&&\\
N11B-4&PGMW 3160 (early KI)&&5&15&0.10\tablenotemark{c}&0.10\\
N180B-1&MGSD 214 (O3-4)&13.4&22&20&0.33&2.97\\
\tableline
\end{tabular}
\tablenotetext{a}{The insufficient spatial coverage of our echelle spectra to the
south of N11B does not allow us to assign accurate boundaries to this expanding shell.}
\tablenotetext{b}{The spectral types of components A--F are taken 
from Walborn et al. (1999)}
\tablenotetext{c}{As the star is evolved, the bubble dynamics must have been 
modified, therefore these numbers may not be meaningful.}
\end{center}
\end{table}
\clearpage

\end{document}